\documentclass[runningheads]{llncs}
\usepackage[T1]{fontenc}
\usepackage{graphicx}
\usepackage{subcaption}
\usepackage{soul}
\usepackage{xcolor}
\usepackage{caption}

\begin{document}
\title{An experimental study of the response time in an edge-cloud continuum with ClusterLink}
%
%
\author{Marc Michalke\inst{1}\orcidID{0000-0003-0766-0397} \and
Fin Gentzen\inst{1}\orcidID{0009-0000-4661-5033} \and
Admela Jukan\inst{1}\orcidID{0000-0002-4434-6340} \and
Kfir Toledo\inst{2}\orcidID{0000-0002-7151-9577} \and
Etai Lev Ran\inst{2}\orcidID{0009-0000-4330-0546}}
\authorrunning{M. Michalke et al.}
%
\institute{Technische Universität Braunschweig, Germany \and IBM Research}
\maketitle              
\begin{abstract}
In this paper, we conduct an experimental study to provide a general sense of the application response time implications that inter-cluster communication experiences at the edge at the example of a specific IoT-edge-cloud contiuum solution from the EU Project ICOS \cite{icos} called ClusterLink \cite{clusterlink-url}. We create an environment to emulate different networking topologies that include multiple cloud or edge sites scenarios, and conduct a set of tests to compare the application response times via ClusterLink to direct communications in relation to node distances and request/response payload size. Our results show that, in an edge context, ClusterLink does not introduce a significant processing overhead to the communication for small payloads as compared to cloud. For higher payloads and on comparably more aged consumer hardware, ClusterLink version 0.2 introduces communication overhead relative to the delay experienced on the link.



\end{abstract}
\captionsetup{aboveskip=0pt,belowskip=0pt}

\section{Introduction}
Over the past decade, edge and cloud computing converged into the so-called continuum, with an expected large spectrum of future service offerings. Edge computing complements traditional cloud computing by moving compute resources closer to the end user at the cost of processing capability. In conjunction with IoT, these technologies are further evolving into the IoT-Edge-Cloud continuum, an ecosystem where the devices with a range of capabilities that jointly process and store data to optimize the user experience. This new computing paradigm however not only differs in computing capabilities of the individual components but also in its connection parameters. For instance, shorter distances toward the end user in the edge are in stark contrast to the comparably higher distances between individual nodes and clusters in the cloud, resulting in proportionate changes in delay, jitter and packet loss along the connections. Additionally, edge nodes might even be connected through wireless and mobile connections, further challenging the performance. As a result, stable application endpoints need to be provided by inter-cluster communication solutions.

Existing research in this regard focuses primarily on optimization of the service selection \cite{bachar23} or models for automated deployment and coordination of applications \cite{ejaz24} while the said aspects of the link characteristics of edge networks or the continuum are less of a focus. On the other hand, previous performance studies with various tools assume cloud environments only, which makes the conclusions not applicable to the network parameters and hardware found in the edge or continuum context. Past work mainly focused on CPU utilization \cite{ejaz24} and throughput \cite{osmani} or round trip times \cite{kannan}. 
In general, studies that show the impact of inter-cluster and inter-node latency on application response times in multi-cluster scenarios at the edge are few and far between, and often assume static node port addressing.

In this paper, we conduct an experimental study to provide a general sense of the application response time that inter-cluster communication experiences at the edge at the example of a specific  compute continuum solution aimed at EU Project ICOS \cite{icos} called ClusterLink \cite{clusterlink-url}. ClusterLink is an open source inter-cluster communication solution, that is used for service to service communication across multiple clouds or remote facilities and follows an open and extensible paradigm\cite{clusterlink}. It was shown to have negligible response times in a data center cloud environment. To experimentally analyze the performance of ClusterLink in a compute contiuum, we create an environment that allows us to manipulate link parameters to emulate different networking topologies like multiple cloud or edge sites. We then conduct a set of tests to compare the application response times via ClusterLink to direct communication in relation to node distances and request/response payload size.
Our contribution can be summarized as follows:
\begin{itemize}
    \item Engineer an experimental software architecture for distributed Kubernetes clusters in an IoT-Edge-Cloud continuum
    \item Experimentally integrating the ClusterLink solution in various setting within an IoT-Edge-Cloud continuum
    \item Measure and analyze response times under different network characteristics and payload sizes.
\end{itemize}

The remainder of this work is structured as follows. Section 2 presents the system architecture of our designed testbed. Section 3 describes the design of the measurement procedure and execution and analyzes the results. Section 4 concludes the paper.


\section{Experimental System Architecture}
Our system architecture aims to replicate a network of distributed Kubernetes nodes that form two distinct clusters where one of them provides a serverless function that should be accessed by an application running on the other cluster. We introduce a Nebula overlay network to solve the connectivity problem at the node level, akin to a real-world edge scenario which includes network gateways with mechanisms like NAT or firewalls, and thus potentially blocking public accessibility of some sites. To solve the problem of inter-cluster connectivity on Kubernetes level, we introduce ClusterLink which allows us to address services of other clusters via their respective Kubernetes DNS names.

\vspace {-0.5cm}
\begin{figure}
\includegraphics[width=\textwidth]{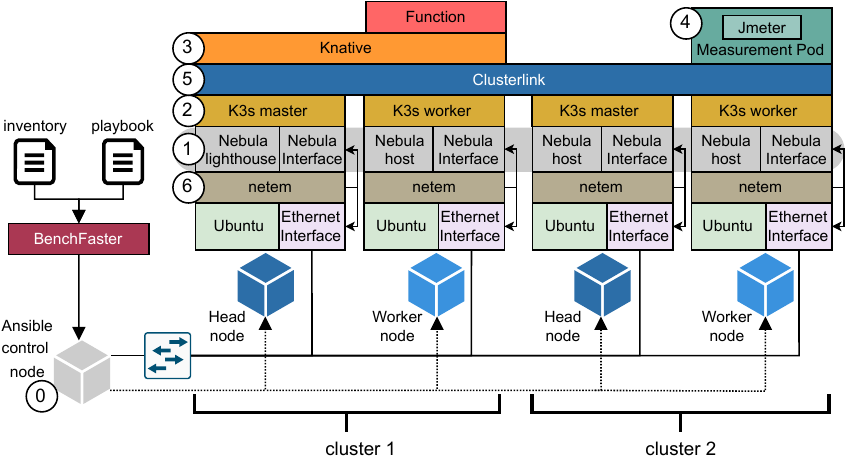}
\caption{Software Stack and Infrastructure} \label{fig:software}
\end{figure}
\vspace {-0.5cm}

For this work, we leverage an enhanced version of an open-source testbed developed in previous work \cite{benchfaas}. It allows for automatic setup of our infrastructure that will be described in detail in this section. 

\subsection{Infrastructure}
As can be seen in Fig. \ref{fig:software}, the setup consists of a set of five different physical machines; two worker nodes and two headnodes that are logically divided into two Kubernetes clusters respectively, and an Ansible control node, as listed in Tab. \ref{tab_hardware}. All physical machines reside in the same LAN and are interconnected in star topology via a Gigabit Ethernet switch. The cluster nodes operate on Ubuntu Server 22.04, while the control node is based on Arch Linux.

To minimize the impact of our diverse choice of hardware platforms on the overall findings, we only compare relative response times instead of conducting measurements of representative, total performance of the solution. The same applies to the choice of Gigabit Ethernet connectivity, since the tests are not designed to exhaust the connection bandwidth.

\begin{table}
\caption{Hardware}\label{tab_hardware}
\begin{tabular}{|l|l|l|l|}
\hline
Cluster & Role & Memory & Processor\\
\hline
None & control & 64GB & Intel(R) Core(TM) i9-10900X CPU @ 3.70GHz \\
1 & headnode & 16GB & Intel(R) Core(TM) i5-7600 CPU @ 3.50GHz \\
1 & worker & 16GB & Intel(R) Core(TM) i5-10600 CPU @ 3.30GHz \\
2 & headnode & 32GB & Intel(R) Core(TM) i5-7600 CPU @ 3.50GHz \\
2 & worker, tester & 8GB & Intel(R) Core(TM) i3-9100 CPU @ 3.60GHz \\
\hline
\end{tabular}
\end{table}

\subsection{Software}\label{sec:software}
Regarding the software, we build upon previous work by extending the open-source benchmarking and validation platform Benchfaster \cite{benchfaster} to allow for automated setup of a Kubernetes-based multi-cluster environment. The framework automatically creates the software stack depicted in Fig. \ref{fig:software} which we descibe below. Our extensions will furthermore be upstreamed to the original open-source project, making our setup reproducible.

\par \textbf{Ansible (0)}
Ansible is an automation software that allows for definition of infrastructure as code through SSH connections, based on sets of instructions, the so-called playbooks \cite{ansible-docs}. Our setup uses a playbook that describes different parts of the system, such as Kubernetes clusters and applications deployed on top of it or the performance tests to be executed, while the machines to be used are defined in an inventory file. All these parts, as well as Ansible itself, in version 2.16.6, which only needs to be deployed in the control node which then provisions all other devices via SSH. Once the setup process is completed and the tests have been executed, the results are transferred to the control node.

\par\textbf{Nebula overlay network (1)}
In a distributed edge network, nodes not only reside behind traditionally expected routers and firewalls administrated and configured by the same entity that deploys the nodes, but might also be located in corporate or private networks behind less permissive gateways or even NAT. Therefore, our testbed deploys Nebula version 1.8.2 \cite{nebula} to create an overlay network, which allows for IP communication between different sites without having to take care of router or firewall configurations. Specifically, the headnode of cluster 1 hosts the Nebula lighthouse which is then used to negotiate direct connections with and between all other nodes and therefore has to be accessible by all of them. In more complex network scenarios, this lighthouse is the only node that has to be publicly accessible without NAT, e.g., in the cloud domain of the continuum. Nebula is used for all inter-cluster communication, while intra-cluster communication is through direct Ethernet connections.

\par\textbf{Kubernetes (2)}
As container orchestrator, we choose k3s, a lightweight Kubernetes distribution designed for resource-limited devices, compatible with most Kubernetes applications (version v1.27.3+k3s1). Each of the k3s clusters is initiated and managed by the respective headnode, which hosts the Kubernetes control plane and is joined by all worker nodes of the same cluster. If a headnode later receives a request for an application service, it will forward the request to a worker running the matching workload. In our setup, the headnode does not host any application containers. The worker nodes on the other hand deploy only application containers as well as the components necessary to enable the management through the headnode.

\par\textbf{Knative (3)}
As an example workload for a distributed environment, we opt for a stateless application based on a serverless paradigm by deploying a test function on top of the serverless platform Knative in version v1.10.0. Test functions can then be deployed based on the needs of the respective test and accessed through one of two ways; directly from within the same cluster through the service name, or from external sources through the Knative Ingress point at port 80 of the respective headnode.

\par\textbf{Measurement pod (4)}
For the scenario of a workload being generated within another cluster, a Kubernetes pod is placed on a worker node cluster 2 based on a modified image of debian trixie-20240423-slim with access to the host filesystem through a volume mount, allowing for loading test definitions and saving the results to it. To prevent impacts from any default resource limitations, the deployment is created with a limit of 2 GB of memory and 2 CPU cores.

\par\textbf{ClusterLink (5)}
ClusterLink is an open-source project that offers a secure and performant solution for interconnecting services across multiple clusters in different domains, networks, and cloud infrastructures. It features fine-grained policy control, allowing for connection management based on various attributes such as cluster, service, or workload levels.
ClusterLink has two main components: (1) Control Plane: Uses Custom Resource Definitions (CRDs) to configure and control the connections between services across multi-cluster environments; and (2) Data Plane: Responsible for establishing secure tunnels for workload-to-service communication between clusters.
It should be noted that while the Nebula overlay network ensures IP connectivity between the two clusters, direct communication between Kubernetes services is still not possible. ClusterLink solves this problem not just by making services available across different domains, networks, and cloud infrastructures, forming a so-called fabric, but also by featuring fine-grained policy control, allowing connection management based on various attributes such as cluster, service, or workload levels \cite{clusterlink-blog}.

\par\textbf{Network Emulator (6)}
Benchfaster can manipulate physical and virtual interfaces to modify link metrics like packet loss, delay, packet corruption or jitter via the tool \textit{netem} which is part of all major Linux distributions. This allows for emulation of different network environments by replicating specific attributes of connections like geographical distance or lossy wireless protocols.

\section{Test Design and Execution}
We assume that the worker nodes in the continuum can dynamically join and leave the continuum, while the headnodes are static. Therefore, there are only two ways to contact a service reliably; either through a control plane that manages these changes dynamically or through the static Knative ingress exposed by the headnode of a cluster. To establish a baseline to which we could compare the ClusterLink response times against, we run the tests directly against this Knative endpoint, i.e., bypassing all ClusterLink components. Then, the same test is executed against the function's internal service exposed through ClusterLink.

Since we replicate the communication between two services, the test assumes two clusters in teh continuum, and targets the serverless function service in cluster 1 and is conducted from a benchmarking pod running on the worker node of cluster 2. This service is reachable via the service name exposed through ClusterLink, or through the Knative ingress provided by the headnode of cluster 1, which we refer to as a direct connection. The serverless functions are based on simple container images with code written in Golang. One, named hello-world, replies to all requests with the string "Hello, World!", providing a minimal payload size to determine potential communication overhead. The second function, named payload-echo, responds to HTTP POST requests by answering with the received payload and, in our case, is provided with a 100KB json document. The benchmarking container is deployed with Jmeter 5.5 installed. It then runs two test definitions as .jmx files against the respective targets, either the service made accessible by ClusterLink or the endpoint exposed by the headnode. These are then repeated across three different network scenarios. 

\subsection{Network Scenarios}\label{sec:endpoints}
\begin{figure}[!ht]
    \centering
    \begin{subfigure}[b]{0.3\textwidth}
        \includegraphics[width=\textwidth]{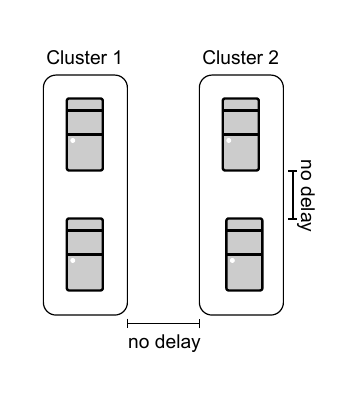}
        \caption{Local}
        \label{fig_scenarios-loc}
    \end{subfigure}
    \hfill
    \begin{subfigure}[b]{0.3\textwidth}
        \includegraphics[width=\textwidth]{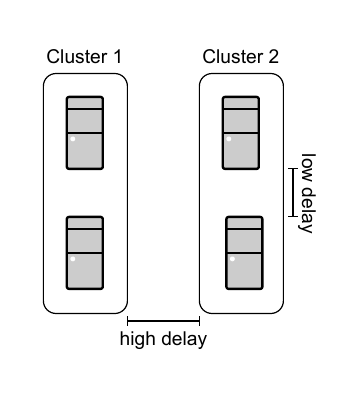}
        \caption{Edge}
        \label{fig_scenarios-edg}
    \end{subfigure}
    \hfill
    \begin{subfigure}[b]{0.3\textwidth}
        \includegraphics[width=\textwidth]{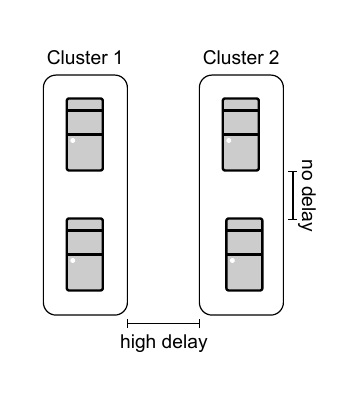}
        \caption{Cloud}
        \label{fig_scenarios-cld}
    \end{subfigure}
    \caption{Network Scenarios}
    \label{fig_scenarios}
\end{figure}

Fig. \ref{fig_scenarios-loc} shows three scenarios analyzed. The local scenario emulates two distinct clusters within the same rack, introducing no additional delay, jitter or packet loss between them. The edge scenario assumes clusters that consist of distributed nodes, while the cluster nodes have double the inter-node distance between them; resulting in double the delay between clusters compared to connections between nodes of the same cluster. The cloud scenario assumes that all nodes of a cluster are concentrated on the same site, while there is a distance to be bridged between clusters, resulting in no latency between nodes but only between clusters.
\vspace {-0.7cm}
\vspace {-0.5cm}
The detailed values are listed in Tab. \ref{tab_netem}, where "phy" represents the values applied to the physical interfaces and therefore all inter-node communication within a cluster and "ovl" the values applied to the virtual Nebula overlay interface, most relevant to the inter-cluster communication. Since the values are applied at all interfaces involved in the communication, this results in, e.g., 10 ms of total delay added to the communication of a node1 of cluster 1 to a node2 in cluster 2 in the edge 10 scenario, since the packet has to cross the physical ($+ 2.5 ms$) and the virtual ($+ 2.5 ms$) interfaces on the source node and then both again on the destination node ($+ 2.5 ms + 2.5 ms$). Similar statements apply for the packet loss and jitter values.
The values for these parameters are derived from the typical network latency reported in \cite{fcc2021} and \cite{opensignal2019} as well as the latency values measured for cloud and edge data centers \cite{charyyev2020} and then scaled up and down accordingly.
\vspace {-0.7cm}
\begin{table}
\caption{Network Scenario Parameters}\label{tab_netem}
\begin{tabular}{|l|l|l|l|l|l|l|}
\hline
Scenario & phy del.(ms) & phy jit.(ms) & phy loss(\%) & ovl del.(ms) & ovl jit.(ms) & ovl loss(\%) \\
\hline 
local       & 0 & 0 & 0 & 0 & 0 & 0 \\
edge 5      & 1.25 & 0.25 & 0.02 & 1.25 & 0.25 & 0.02 \\
edge 10     & 2.5 & 0.5 & 0.04 & 2.5 & 0.5 & 0.04 \\
edge 20     & 5 & 1 & 0.08 & 5 & 1 & 0.08 \\
edge 30     & 7.5 & 1.5 & 0.12 & 7.5 & 1.5 & 0.12 \\
cloud 5     & 0 & 0 & 0 & 2.5 & 0.5 & 0.04 \\
cloud 10    & 0 & 0 & 0 & 5 & 1 & 0.08 \\
cloud 20    & 0 & 0 & 0 & 10 & 2 & 0.16 \\
cloud 30    & 0 & 0 & 0 & 15 & 3 & 0.24 \\
\hline
\end{tabular}
\end{table}
\vspace {-0.7cm}

\subsection{Test Execution of Response Time}
Each test consists of 20 successive HTTP sent to the respective endpoint (hello-world direct, hello-world ClusterLink, payload-echo direct, payload-echo ClusterLink) from a single thread, all reusing the same TCP connection with keep-alive enabled in Jmeter as well. For the hello-world function, HTTP GET requests are sent, while the HTTP POST requests to the payload-echo function are provided with a request body of 1KB, 10KB, 100KB and 1MB. Both tests are intentionally designed not to stress the system but instead to measure the response time of these requests.

We repeat this set of tests and network scenarios a total of ten times against a clean environment to ensure reproducibility. The cleanup, i.e., the removal of the installed software and reset of the interface configuration, as well as recreation of the environment is provided by the Benchfaster framework.


\subsection{Measurements}
\begin{figure}[!ht]
    \centering
        \includegraphics[width=\textwidth]{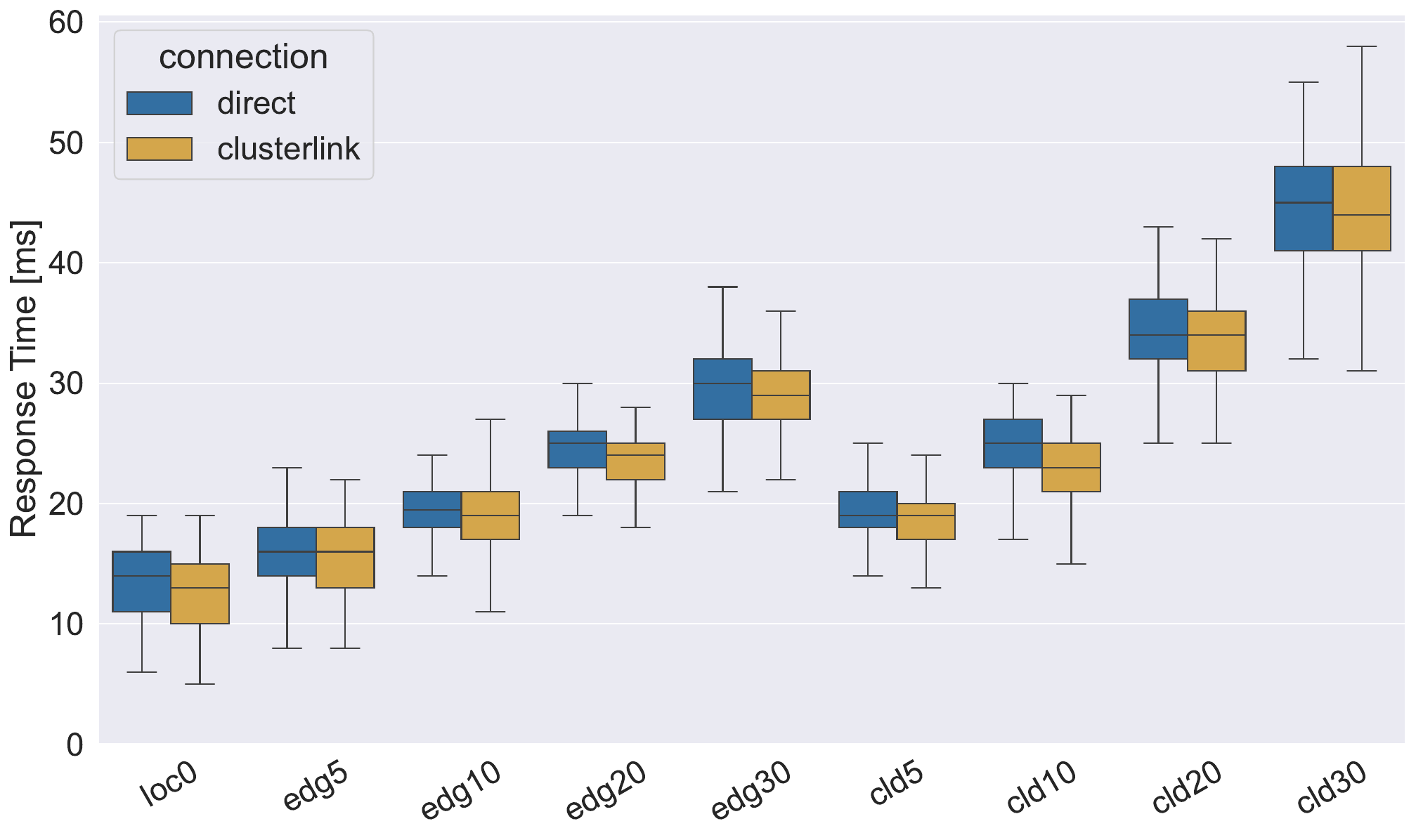}
    \caption{Hello-World Response Time}
    \label{fig_res_helloworld}
\end{figure}
Fig. \ref{fig_res_helloworld} shows the results of the hello-world test. We use the local network scenario to determine the default application response time for this function on top of Kubernetes and Knative in a local area network scenario where network delay is negligible, therefore establishing a baseline of around 10 ms.
It can be observed that the differences in application response time between ClusterLink and the direct connection do not exceed the expected margin of error. Since the connections are reused, ClusterLink only adds two round-trip-times for the initial session establishment, resulting in the very first response time of each test being three times the value of the later requests. During this session setup, the cross-cluster access policies are checked and encryption keys are negotiated for securing connections leaving the cluster. Once the session is established, ClusterLink's impact on latency is small and bounded (<1ms) for small payloads, unnoticeable in WAN, wireless and other networks that exhibit natively high delays; and applications that have non-negligible request-response processing times.
With increasing delay, the values for direct and ClusterLink connections then increase linearly based on the introduced total amount of delay for the connection without significant differences.

\vspace{-0.5 cm}
\begin{table}
\caption{Payload Test Results for direct (D.) and ClusterLink (C.)}\label{tab_results}
\begin{tabular}{|l|l|l|l|l|l|l|l|l|}
\hline
Scenario & C. 1KB & C. 10KB & C. 100KB & C. 1MB & D. 1KB & D. 10KB & D. 100KB & D. 1MB \\
\hline 
local       & 15.63 & 28.02 & 63.28 & 231.10 & 16.30 & 21.02 & 62.11 & 223.49 \\
edge 5      & 18.86 & 34.73 & 73.47 & 267.58 & 18.63 & 25.52 & 68.42 & 237.29 \\
edge 10     & 23.18 & 43.62 & 89.1 & 343.10 & 22.80 & 32.36 & 78.52 & 336.38 \\
edge 20     & 28.17 & 92.29 & 116.32 & 372.92 & 28.09 & 41.43 & 93.75 & 363.04 \\
edge 30     & 38.15 & 95.26 & 137.67 & 499.91 & 33.64 & 49.24 & 121.54 & 405.33 \\
cloud 5     & 21.93 & 44.02 & 88.48 & 352.14 & 23.0 & 34.04 & 78.95 & 344.92 \\
cloud 10    & 27.55 & 57.02 & 111.84 & 396.10 & 27.83 & 39.49 & 93.12 & 359.35 \\
cloud 20    & 44.76 & 81.88 & 174.74 & 561.19 & 38.33 & 59.32 & 143.19 & 528.92 \\
cloud 30    & 100.77 & 105.04 & 263.03 & 759.98 & 47.73 & 80.14 & 201.56 & 690.64\\
\hline
\end{tabular}
\end{table}
\vspace{-0.5 cm}

\begin{figure}[!ht]
    \centering
    \includegraphics[width=\textwidth]{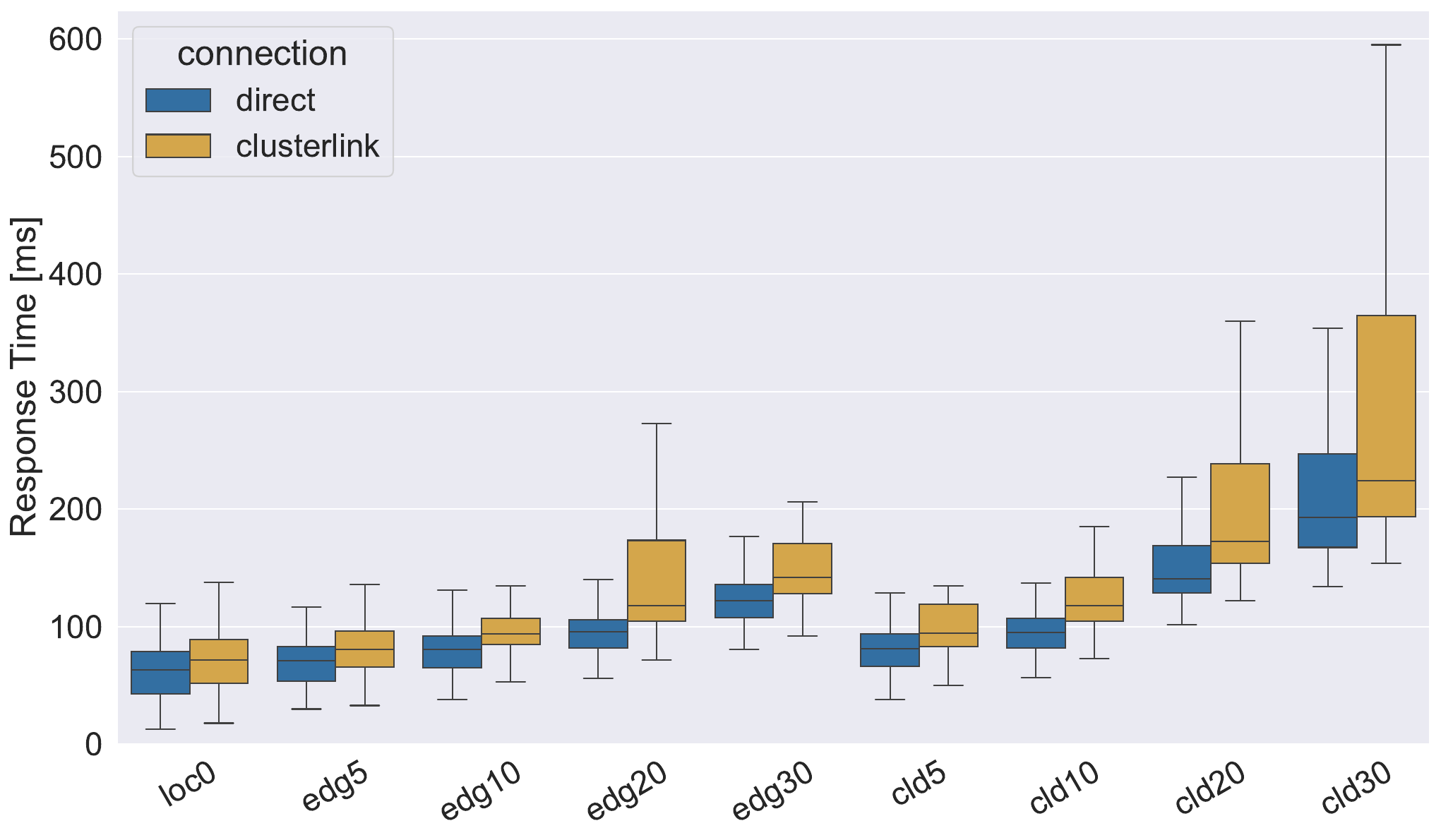}
    \caption{Payload 100KB}
    \label{fig_res_payload100}
\end{figure}

Regarding the payload test, since the results of the 1KB test do not significantly differ from the hello-world results while the trends observed for 10KB and 1MB similar to larger payloads, we show the measurements for 100KB payload in Fig. \ref{fig_res_payload100}, while the IQR mean values of all measurements can be found in Tab. \ref{tab_results}. The main trend we can observe for ClusterLink connections with higher payloads is an increased impact of the inter-node latency on the overall response times than on direct connections. This effect is barely noticeable in our local scenario, with its relevance being proportional to the highest delay experienced on inter-node and inter-cluster connections and only when transporting high payloads. This can potentially be attributed in parts to the TCP connections used by ClusterLink and therefore the utilized buffers filling up, leading to additional delays in the transmission into both directions. Another factor might be the encryption utilized by ClusterLink, which could introduce processing overhead when encrypting and decrypting the data before and after transmission. The latter however can be expected to have a much smaller effect on more capable or modern hardware platforms, especially if they leverage dedicated cryptoprocessors.

\vspace{-0.5 cm}
\vspace{-1.5 cm}

\section{Conclusions}
We experimentally analyzed and compared the inter-cluster connectivity solution ClusterLink to static endpoints hosted on the headnode of each cluster in an edge-cloud continuum. Our measurmeents show that, even on comparatively more aged consumer hardware, ClusterLink does not introduce a significant processing overhead to the communication for small payloads. While the checking of the cross-cluster policies and negotiation of encryption keys introduces two additional round-trip times for each session establishment, ClusterLink's impact on latency afterwards is small and bounded (<1ms) for small payloads, unnoticeable in WAN, wireless and other networks that exhibit higher delays; and applications that have non-negligible request-response processing times.  For bigger payloads, ClusterLink, in its current version and on the tested hardware, does impact the overall response time in relation to the delay experienced on the link. Since this software is still in early development and did not yet hit a stable release target at the time of writing, this might also be subject to change.

Future tests should be repeated across a wider variety of hardware selections. The introduction of our Nebula overlay network has potentially increased the communication overhead and therefore should ideally be rendered obsolete by other mechanisms for inter-cluster connectivity. Also, a comparison to established inter-cluster communication networks would be valuable to highlight specific and general up- and downsides of inter-cluster communication solutions.

\begin{credits}
\subsubsection{\ackname}This work was partially supported by the project ”Towards
a functional continuum operating system (ICOS)” funded by
the European Commission under Project code/Grant Number
101070177 through the HORIZON 2020 program..
\end{credits}
%
%
%
%

\end{document}